\documentclass [11pt,twoside]{article}

\usepackage{shrthnds}
\usepackage{cite}
\usepackage{graphicx,amsmath}
\usepackage{color}
\usepackage{subfigure}
\usepackage{setspace}

\usepackage{multirow}

\usepackage[hang,small]{caption}

\usepackage{geometry}
\usepackage{fancyhdr}
\usepackage{titlesec,titletoc}
  \titleformat{\section}{\Large\sf\bfseries}{\thesection}{1em}{}
  \titleformat{\subsection}{\large\sf\bfseries}{\thesubsection}{1em}{}

\title{\sf\bfseries 
The fine tuning of the cosmological constant in a conformal model
}

\author{\normalsize  
Pankaj Jain\footnote{email: pkjain@iitk.ac.in}~,
 Gopal Kashyap\footnote{email: gopal@iitk.ac.in}
and Subhadip Mitra\footnote{email: subhadipmitra@gmail.com}
}
\date{}%{\today}

\begin{document}
\maketitle
\vspace{-0.6cm}
\bc
{\small Department of Physics, IIT Kanpur, Kanpur 208 016, India}
\ec

\centerline{\small\date{\today}}
\vspace{0.5cm}

\bc
\begin{minipage}{0.9\textwidth}\begin{spacing}{1}{\small {\bf Abstract:}
%%%%%%%%%%%%%%%%%%%%%%%%%%%%%%%%%%%%%%%%%%%%%%%%%%%%%%
% ABSTRACT
%%%%%%%%%%%%%%%%%%%%%%%%%%%%%%%%%%%%%%%%%%%%%%%%%%%%%%
We consider a conformal model involving two real scalar fields in which the
 conformal symmetry is broken by
a soft mechanism and is not anomalous. One of these scalar fields is
representative of the standard model Higgs. The model predicts exactly zero 
cosmological constant. In the simplest version of the model, 
some of the couplings need to be fine tuned to 
very small values. 
We formulate the problem of fine tuning of these 
couplings. We argue that the problem arises since we require a soft
mechanism to break conformal symmetry. The symmetry breaking 
is possible only if the
scalar fields do not evolve significantly over the time scale of
the Universe. We present two solutions to this fine tuning problem.
We argue that the problem is solved if the classical value of
one of the scalar fields is super-Planckian, i.e. takes a value much
larger than the Planck mass. The second solution involves introduction
of a strongly coupled hidden sector that we call hypercolor. In this case
the conformal invariance is broken dynamically and triggers the
breakdown of the electroweak symmetry. 
We argue that our analysis applies also to the case of the standard model Higgs
multiplet. 

%%%%%%%%%%%%%%%%%%%%%%%%%%%%%%%%%%%%%%%%%%%%%%%%%%%%%%
}\end{spacing}\end{minipage}\ec

\vspace{0.5cm}\begin{spacing}{1.1}

%%%%%%%%%%%%%%%%%%%%%%%%%%%%%%%%%%%%%%%%%%%%%%%%%%%%%%
% MAIN CONTENT STARTS HERE
%%%%%%%%%%%%%%%%%%%%%%%%%%%%%%%%%%%%%%%%%%%%%%%%%%%%%%
\section{Introduction}
The idea that conformal invariance \cite{Weyl:1929,Deser,Dirac1973,Sen1971,Utiyama:1973,Utiyama:1974,Freund:1974,Englert:1976,Hayashi:1976,Hayashi:1978,Nishioka:1985,Padmanabhan85,Ranganathan:1987,Cheng88,thooft14}  might solve the problem of fine tuning
of the cosmological constant \cite{Weinberg,Padmanabhan} 
is very old \cite{Wetterich,Mannheim89} and has
attracted considerable interest in the literature 
\cite{JM,JMS,Shaposhnikov:2008a,Shaposhnikov:2008b,JM09,Nishino09,JM10_2,JMPS,Mannheim09,rajpoot3,bars1,Jain14}. 
A theory with conformal invariance does not permit a cosmological constant
and hence might impose some constraint on its value. However due to 
conformal anomaly it is not clear that it is possible to maintain a small
value of the cosmological constant at loop orders even if the action
displays classical conformal invariance.   
 Furthermore one requires some
source of dark energy\cite{Ratra,Peebles,Copeland}. Hence the model has to provide its very small value 
without fine tuning.  

It has been shown that conformal invariance can be implemented in the 
full quantum theory if we use a dynamical scale for regularization
\cite{Englert:1976,JM,JMS,Shaposhnikov:2008a,Shaposhnikov:2008b}.
This is implemented by introducing a real scalar field in the model. 
 The
procedure has been called the GR-SI prescription in \cite{Shaposhnikov:2008a}. 
In this case the conformal symmetry is broken by a soft mechanism. It
may be spontaneously broken \cite{Englert:1976,Shaposhnikov:2008a,Shaposhnikov:2008b} or broken by the background cosmic evolution \cite{JM,JMS}. 
This leads to a non-zero classical value of the real scalar field which
provides a scale for regularization. 
One finds that the implications of conformal symmetry are maintained
even in the full quantum theory. However the theory predicts  
renormalization group evolution of the coupling
parameters despite being conformally invariant \cite{Shaposhnikov:2008a}.

 The perturbation theory in the GR-SI prescription becomes
more complicated involving additional scalar interaction terms.  
It has been argued \cite{Tkachov09,Jain10a,Jain14} that 
these additional terms may make the model non-renormalizable. However 
so far it has not been explicitly shown that this is true. In any case, 
the additional terms are suppressed by Planck mass and hence the
problem is not more severe in comparison to the non-renormalizability
of gravity \cite{Tkachov09,Jain10a,Jain14}. 
Furthermore these additional terms are practically irrelevant if we ignore 
contributions due to the added real scalar field, denoted by 
$\chi$ in this paper. Hence if we confine
ourselves to a study of only the standard model fields, we recover the 
standard perturbation theory.    

Another problem with the model is that one of the allowed terms in the
action has to be set to zero at each order in the perturbation theory. 
This is required in order to break conformal invariance spontaneously
\cite{Shaposhnikov:2008a}. Else it is not possible to implement the 
GR-SI prescription. Alternatively one needs to maintain a very small 
value of the coupling constant corresponding to
this additional term at each order in perturbation theory
\cite{JMPS,JMPRD}. The problem is again traced to the small value of
the cosmological constant in comparison to other scales in the theory
and hence the problem is not solved. 

In a recent paper \cite{Jain14} 
we have shown that the fine tuning problem of cosmological constant
gets partially resolved if we add small conformal symmetry breaking terms
to the action. In this case we still demand that one of the terms
in the conformal action is identically equal to zero despite it being
allowed by the symmetry of the theory. Once this term is set to zero
the conformal action predicts identically zero cosmological constant. 
We can add small symmetry breaking terms. The small values of these 
terms are preserved in perturbation theory since they receive zero
contribution from the conformal action. We have shown in \cite{Jain14}
that these symmetry breaking terms lead to the observed dark energy.

In the present paper we carefully formulate the problem
associated with the fine tuning of
the cosmological constant within the conformal model. We argue that
since the model displays exact conformal invariance even for the
dimensionally regulated action, we expect the trace of the energy
momentum tensor, $T^\mu_\mu$, to be equal to
zero. Still, one has to choose the gravitational action carefully
in order that the curvature scalar $R$ is zero. 
This is true only for a special choice of gravitational
action \cite{CCJ}. In particular we do not impose the requirement
that the gravitational action must also be invariant under conformal
transformation. Our gravitational action may also be obtained by 
requiring local conformal invariance and imposing a particular gauge choice
\cite{Padmanabhan85}. In any case, for a wide range of gravitational
actions, we find that $R$ is zero as long as we ignore the 
cosmological evolution in computing the matter contributions. 
So the value of these contributions is controlled by the background
Hubble parameter and is necessarily very small.
Hence the problem of fine tuning of the cosmological constant in these
models does not manifest itself in the value of $T^\mu_\mu$ or $R$
but in the need to have a soft mechanism to break conformal symmetry. 

In the simplest version of the model, the conformal symmetry is
broken spontaneously. This is possible only if one of the parameters in the
theory, which we denote as $\lambda$ in our paper, is taken to be
zero or very very tiny. This parameter also gets large corrections 
at loop orders and has to be fine tuned order by order in perturbation
theory \cite{Tamarit13}. 
If this parameter is not fine tuned then we find that the 
scalar fields quickly decay to zero and the conformal symmetry is
restored. We propose two solutions to this problem. We show that the
problem of fine tuning at loop orders is absent if the classical 
value of the scalar field $\chi$ is taken to be super-Planckian,
i.e. much larger than the Planck mass. The second solution, which
appears to be more interesting, involves the introduction of a 
strongly coupled sector, such as technicolor \cite{Farhi}. The generation of
condensates in this sector breaks conformal symmetry. We choose a
specific model in which the strongly coupled sector acts as a dark
sector \cite{Meissner07,Meissner08,Hur11} and couples to the standard 
model fields only through the scalar
field $\chi$. We show that in such models the problem of fine tuning 
of the cosmological constant is absent.  
Our proposal is applicable both in the case of global or local 
conformal invariance, although in the present paper we confine our
discussion to global invariance. We point out that in this paper 
we shall be primarily concerned with solving the fine tuning problem 
associated with the cosmological constant. We shall not address the
issue of the source of dark energy and the resulting cosmological 
evolution. However once we are able to construct a model in which 
 the cosmological constant is naturally zero or very small, the issue
of dark energy can be addressed systematically. Furthermore the model
does admit a dark sector which can, in principle,
 lead to the observed dark components. 

Before we proceed further, we note that the
 implications of local conformal invariance
have been extensively studied in literature \cite{Huang1989,Hochberg91,Wood92,Wheeler98,Feoli98,Pawlowski99,Demir2004,Wei2006,Moon,Maki12,Codello,GK,Quiros:2014,Dengiz14,Padilla,GK2014,Salvio14}. The implications of global scale invariance have also been 
investigated
\cite{GarciaBellido:2011de,GarciaBellido:2012zu,Jain13,Armillis,Henz,Gorbanov1,Gretsch,Khoze}.
Other proposals to solve the cosmological constant problem are discussed in
\cite{Weinberg,Aurilia,VanDer,Henneaux,Brown,Buchmuller,Buchmuller1,Henneaux89,Sorkin,Sahni1,Sundrum,Diakonos09,Barrow11,Barrow11a,Demir11}.

\section{Fine Tuning Problem in a Conformal Model}
\label{sec:problem}
Let us first consider a simple model invariant under global conformal 
invariance containing two
real scalar fields, $\chi$ and
$\phi$. 
The action may be written as,  
\begin{equation}
{\mc S} = {\mc S}_G + \int d^4x \sqrt{-g} 
\left[ 
{1\over 2}g^{\mu\nu} \partial_\mu\chi \partial_\nu\chi
+{1\over 2}g^{\mu\nu} \partial_\mu\phi \partial_\nu\phi
-V(\chi,\phi)\right]
\label{eq:SC}
\end{equation}
where $\mc S_G$ is the gravitational action and the potential $V$ is given by,
\begin{equation}
V(\chi,\phi) =  {\lambda\over 4}\chi^4+{\lambda_1\over 4}\left(\phi^2-\lambda_2\chi^2
\right)^2
\label{eq:Vchiphi}
\end{equation}
Here $\lambda$, $\lambda_1$ and $\lambda_2$ are
coupling parameters. 
The model displays invariance under the transformation, 
\begin{equation}
g_{\mu\nu}\rightarrow \Omega^2 g_{\mu\nu}\, , \chi\rightarrow 
{\chi\over \Omega}\, , \phi\rightarrow 
{\phi\over \Omega} 
\label{eq:CT}
\end{equation}
where $\Omega$ is the transformation parameter. 
Due to conformal symmetry we do not have a cosmological constant 
term in the action. It is generally expected that the conformal symmetry
is anomalous and hence the theory will necessarily lead to a large
cosmological constant. However, as explained in the introduction,
here we shall use the
 GR-SI prescription   
  \cite{Shaposhnikov:2008a} for 
regularization that does not break conformal invariance\cite{Englert:1976,JM,JMS,Shaposhnikov:2008a,Shaposhnikov:2008b}. 
With this procedure, the theory preserves conformal invariance at 
 all orders in perturbation theory and does not generate a non-zero
cosmological constant even at loop orders.

We are interested in identifying $\phi$ with the Higgs and hence we expect
its vacuum expectation value (VEV) to be equal to the electroweak symmetry breaking scale.
The gravitational action may be taken of the form,
\begin{equation}
S_G = \int d^4x\sqrt{-g} \left( {1\over 8}\zeta^2 R  \right)
\label{eq:SG1}
\end{equation}
where $R$ is the Ricci scalar,  
\begin{equation}
\zeta^2 =\beta\chi^2+\beta_1\phi^2 
\label{eq:zeta}
\end{equation}
and $\beta$, $\beta_1$ are parameters.  
As we shall see later, this is not the most convenient form of the
gravitational action for our purpose. We, however, display it here since it is 
invariant under conformal transformation, Eq. \ref{eq:CT}, and also 
often found in literature (see for example, 
\cite{Cheng88,JMS,Shaposhnikov:2008a,Shaposhnikov:2008b}). 
Assuming that the parameters $\beta$ and 
$\beta_1$ are of order unity, we expect the vacuum expectation value of
$\chi$ to be of the order of the Planck mass, $M_{PL}$, in order to reproduce the
observed value of the gravitational constant. 
As we shall see later this may not be a good motivation for setting
$\langle\chi\rangle$ equal to $M_{PL}$.
However this may be a good choice for other reasons. In any case,
the precise value of $\langle\chi\rangle$ does not have any effect on the fine 
tuning problem under consideration.   

We next determine the minimum of the potential. It has been shown that
a minimum appears at non-zero values of the fields 
only if we set $\lambda=0$ \cite{Shaposhnikov:2008a}. In that case the potential
is minimized for,  
\begin{equation}
v = \sqrt{\lambda_2}\eta
\label{eq:potentialmin}
\end{equation}
where $v$ and $\eta$ represent the classical values of the fields
$\phi$ and $\chi$ respectively. 
We expect that $v$ is of the order of the
 electroweak symmetry breaking scale and $\eta$
of order $M_{PL}$. Hence the parameter $\lambda_2\sim 10^{-34}$.  
This parameter is small but by itself does not lead to any fine tuning
problems, as we shall show below. The main problem arises since we need
to set $\lambda=0$ without invoking any symmetry. 

We point out that we need not necessarily demand that the potential
be minimized. For example, we might consider a very small value of
the parameter $\lambda$. In this case the minimum of the potential
occurs at $\chi=\phi=0$. 
However we can choose initial conditions such that  
  the fields are initially at some non-zero
values and evolve with time towards the minimum of the potential. 
If the field $\chi$ takes a value close to $M_{PL}$ today then the parameter
$\lambda$ has to be order $10^{-120}$. 
For such small values of this parameter we expect that the fields
will evolve very slowly towards the minimum and satisfy the standard 
slow roll conditions. Hence the small non-zero value of the potential
acts as dark energy with the equation of state parameter $w$ very close
to $-1$ and we generate an effective cosmological constant in our model. 
The parameter $\lambda$ is the source
of the fine tuning problem and reminiscent of the standard cosmological
constant problem. Hence despite the absence of the cosmological constant
in this model, its fine tuning problem is still present. 

We clarify that in this paper we are not concerned with 
an explanation of the very small parameters in the action. Hence we shall
allow small parameters in the tree level action. However these parameters
may acquire large corrections at loop orders and hence may have to be
fine tuned order by order in perturbation theory. Such a problem does
appear in the model under consideration and we shall focus on 
trying to find a solution to this problem. 

Before proceeding further we point out that we may also seek a dynamical
solution which involves a cosmological evolution assuming an FRW metric.
A solution of this type can be found in this two scalar field model  
\cite{JMS}. This also leads to a soft breaking of the conformal symmetry
with constant $\phi$ and $\chi$.
In this case also the fine tuning problem remains since the coupling
$\lambda$ has to be tuned to very small values.

We next explain how the fine tuning problem appears order by order in
perturbation theory. Before considering the conformal model, let us
review how fine tuning of cosmological constant appears in the standard 
model within the framework of dimensional regularization. 
The problem of fine tuning of the cosmological constant is often
illustrated with an explicit ultraviolet cut-off regulator. But
obviously, even within the dimensional regularization scheme where
no such cut-off appears, the problem remains.
Let $\Lambda$
be the cosmological constant, which has to take a very small value.
As we compute quantum corrections to the vacuum energy density we 
expect to get divergent terms proportional to $v^4/\epsilon$ besides finite
terms of order $v^4$. Here $v$ is the scale of electroweak symmetry 
breaking and $\epsilon=4-d$. In order to remove the divergence we need
to add a cosmological constant counter term. However this leaves behind
a finite term of order $v^4$ which is very large in comparison to the
observed value of $\Lambda$. Hence the finite part of the counter term
must be chosen in order to cancel this term very precisely. This is
the source of the fine tuning problem of the cosmological constant
within the framework of dimensional regularization. 

We now return to the corresponding problem in our conformal model.
The fact that problems are likely to appear is
obvious from the form of the potential. Let us reexpress the potential
in the following form 
\begin{equation}
V(\chi,\phi) = {1\over 4} \left(\lambda_1\lambda_2^2+\lambda\right)\chi^4
+{\lambda_1\over 4}\phi^4-{1\over 2}\lambda_1\lambda_2\phi^2\chi^2
\end{equation}
The coupling $(\lambda_1\lambda_2^2+\lambda)$ of the $\chi^4$ term is
the source of problem. We find that $\lambda\ll\lambda_1\lambda_2^2$.
At loop order the coupling $\lambda$ receives contributions which are
very large. These have to  
be fine tuned order by order in perturbation theory in order to maintain
the extremely tiny value of $\lambda$. The fact that such terms appear
is clear from the calculation of the effective potential at one loop
\cite{Shaposhnikov:2008a}. One finds that the counter terms are such
that the order of magnitude of their finite parts is much larger in
comparison to $\lambda$.  

Besides the contributions from the potential there also exist other
hidden contributions which arise due to the special structure of the
GR-SI prescription. 
To see these, let us consider a few relevant terms from the  conformal action in $d$ dimensions,
\begin{equation}
S_d = \int d^dx\sqrt{-g}\left[ {1\over 8}\zeta^2 R + 
{1\over 2} g^{\mu\nu} \partial_\mu\chi \partial_\nu\chi -
{1\over 4} g^{\mu\rho} g^{\nu\sigma}{\cal E}_{\mu\nu} {\cal E}_{\rho\sigma}
(\zeta^2)^\delta
- {1\over 4}\lambda\chi^4(\zeta^2)^{-\delta} + ...\right]
\label{eq:actionConformal_d}
\end{equation}
where 
$\delta = (d-4)/(d-2)$ and
${\cal E}_{\mu\nu}$ generically represents the field tensor for a gauge field. 
 The key point of the GR-SI regularization
is the appearance of terms like $(\zeta^2)^\delta$. Such terms 
vanish in four dimensions and are included in order to make the action
conformally invariant in $d$ dimensions. The transformation of different
fields in $d$ dimensions is given by, 
\begin{equation}
\chi\rightarrow {\chi\over \Omega}\, ,g_{\mu\nu}\rightarrow
\Omega^{\,b} g_{\mu\nu}\,, 
 A_\mu\rightarrow A_\mu\,, \Psi\rightarrow \Psi/\Omega^{\,c}
\label{eq:globalconftrans}
\end{equation} 
where $b=4/(d-2)$,  $c=(d-1)/(d-2)$, $\Psi$ is a fermion field and $A_\mu$
a vector field. One can easily check that the action,  
Eq. \ref{eq:actionConformal_d}, is invariant under this transformation. 
We point out that we have used the field $\zeta^2$ in both the vector 
field kinetic energy term and the scalar field potential term. In
principle one could introduce a different combination of $\chi$ and
$\phi$ in these terms and hence introduce additional parameters. For 
simplicity here we shall assume that the field used in all these terms
is the same as defined in Eq. \ref{eq:zeta}. 

We obtain a well defined perturbative expansion in the GR-SI prescription
as long as the field $\chi$ and $\phi$ acquire non-zero values classically.
This would imply that the classical value of $\zeta$ is,
\begin{equation}
\zeta^2_{cl} = \beta \eta^2+\beta_1 v^2
\label{eq:zetacl}
\end{equation} 
where $\eta$ and $v$ are the classical values of the fields $\chi$ and
$\phi$ respectively. Let us consider the term $(\zeta^2)^\epsilon$ where
$\epsilon$ is a small parameter which goes to zero as $d\rightarrow 4$. 
We have,
\begin{eqnarray}
(\zeta^2)^\epsilon &=& \exp\left(\epsilon\log(\zeta^2)  \right)\nonumber\\
&=& 1+\epsilon\log(\zeta^2_{cl}) + \epsilon \log\left[1+2{\beta \eta\hat \chi
+\beta_1 v\hat\phi\over \zeta^2_{cl}} + {\beta \hat \chi^2
+\beta_1 \hat\phi^2\over \zeta^2_{cl}}  \right]+...
\label{eq:expepsi}
\end{eqnarray}
where we have expanded the fields $\chi$ and $\phi$ around their classical 
values, such that,
\begin{eqnarray}
\chi = \eta + \hat \chi\nonumber\\
\phi = v + \hat \phi
\end{eqnarray}
and kept only the leading order term in $\epsilon$. 
We can now expand the second log term on the right hand side of 
Eq. \ref{eq:expepsi}. This will generate an infinite series of terms
powers of the fields $\hat \chi$ and $\hat \phi$. If we assume that
$\eta\sim M_{PL}$, these terms will be suppressed by Planck mass. However
these terms also lead to corrections to the Green's functions which involve
external $\hat \chi$ legs of same order as those given by the potential terms. 
We discuss these terms explicitly in section \ref{sec:oneloop}.  
The presence of these terms makes the problem of fine tuning 
even more serious. In particular these terms imply that the problem is 
present even in the composite Higgs models in which the potential terms
proportional to the parameter $\lambda_1$
(see Eq. \ref{eq:Vchiphi}) are absent. 

An important issue within the framework of the conformal regularization
is that the resulting theory may not be renormalizable
\cite{Tkachov09,Jain10a,Jain14}. 
This is because of
the additional terms generated by the conformal regulator as displayed in
Eq. \ref{eq:expepsi}. However, as mentioned in the Introduction,
 so far there does not exist a systematic 
proof in the literature that the resulting theory is indeed not renormalizable.
It seems to us the full theory has not been properly analysed
in the literature so far and we can only regard this as a speculation.
We point out that there exist additional parameters in the theory,
such as $\beta$, $\beta_1$ and it is possible that some of the additonal
divergences can be absorbed in these parameters. In any case 
these additional contributions are suppressed by Planck mass
and hence yield corrections of the order of quantum gravity. We,
therefore, argue that even if the problem is present, it
 is only as severe as that of 
non-renormalizability of gravity \cite{Jain10a,Jain14}.  

\subsection{Gravitational Action}
\label{sec:gravity}
A possible choice of gravitational 
action is displayed in Eq. \ref{eq:SG1}. With this choice the entire action is 
invariant under conformal transformations, Eq. \ref{eq:CT}. 
In this case we find 
that the Ricci scalar, $R$, is not zero and gets 
contributions from several terms in the matter action whose VEVs 
are not zero. These effectively act as the
cosmological constant. Hence,  
for our purpose this choice of gravitational action is not very convenient. 
We point out that several terms in the action, 
including the kinetic energy terms of gauge fields, 
 contribute at loop orders due to their coupling to the scalar field 
in dimensions different from 4.  
Hence despite invoking exact, nonanomalous, conformal invariance
we will predict a rapid exponential cosmological expansion due to vacuum 
contributions unless the different
terms miraculously cancel against one another. 

Alternatively we may choose the standard gravitational Lagrangian, which is
proportional to $M_{PL}^2 R$. In this case the Einstein's equations imply
that $R$ is proportional to the trace of the energy momentum tensor, $T^\mu_\mu$. We expect this trace to be zero in our case due to exact conformal symmetry.
However in the presence of scalar fields, the trace is found to be
proportional to total derivative terms, such as,
\begin{equation}
 \partial_\mu\left(\chi\partial^\mu\chi\right)
\label{eq:totalderivative}
\end{equation}
The vacuum expectation values of such terms vanish in the case of flat 
space-time. In the case of an expanding Universe, these would be 
proportional to the derivatives of the metric and we expect that these
would be small. Such terms are absent if we choose the gravitational
action to be of the form,
\begin{equation}
S_G = \int d^4x\sqrt{-g} \left( {M_{PL}^2\over 16\pi} R -\xi\chi^2 R  \right)
\label{eq:SG2}
\end{equation}
where the first term is the standard Einstein action and second term
is the conformal coupling introduced in Ref. \cite{CCJ}. In $d$ dimensions
the parameter, 
\begin{equation}
\xi= {(d-2)\over 4(d-1)}\, ,
\label{eq:confcoup1}
\end{equation}
which is equal to $1/6$ for $d=4$ \cite{CCJ} (see also \cite{Wehus}). In this special case we find that
$R\propto T^\mu_\mu$ is identically equal to zero in $d$ dimensions. 
In Eq. \ref{eq:SG2} we have displayed the gravitational action
for one scalar field. It can be suitably generalized in the 
presence of additional scalar fields.
This action is clearly elegant and appealing 
but not absolutely required for our purpose since the vacuum values of
the extra terms are either zero or very small.   
Hence we need not make the choice for $\xi$ given in Eq. \ref{eq:confcoup1}.

If we impose local conformal invariance then
the model with $\xi=0$ may also be obtained by choosing a particular 
gauge in which $\chi$ is set equal to a constant \cite{Padmanabhan85}.  
Hence this model may simply represent a particular gauge choice
rather than explicit conformal breaking. In the present paper 
we will confine ourselves mostly to the case of global conformal
symmetry. However we expect that our results should apply also
 for the case of local conformal invariance.

We emphasize that due to conformal invariance we find that in $d$ dimensions,
either
\begin{equation}
T^\mu_\mu = 0
\end{equation}
for $\xi$ given by Eq. \ref{eq:confcoup1} or is proportional to terms such
as given in Eq. \ref{eq:totalderivative}
for other values of $\xi$. 
In the case of classical conformal invariance, such an equation follows
only as a formal statement corresponding to the unregulated action. 
Regularization introduces a mass scale in the action which breaks
conformal invariance \cite{Delbourgo}. 
Hence in this case we cannot trust its consequences at loop orders. However
in the present case, this equation follows 
even for the regulated action. Hence this will lead to zero contributions
to vacuum energy at all orders in perturbation series. 
We also point
out that in establishing this relationship we need to use the
equations of motion in $d$ dimensions. In quantum theory these equations
are interpreted as the Heisenberg operator equations \cite{Nishijima}
 and since they are derived
from the regulated action, we should be able to trust their implications. 
Furthermore, the basic point that we can trust the consequences of conformal 
invariance in the GR-SI prescription has already been made in
Refs. \cite{Englert:1976,JM,JMS,Shaposhnikov:2008a,Shaposhnikov:2008b}.

The fact that $T^\mu_\mu=0$ in $d$ dimensions implies that its VEV 
 also vanishes identically. Alternatively if $\xi$ takes
a value different from that given in Eq. \ref{eq:confcoup1},
 it is proportional to a total derivative term,
displayed in Eq. \ref{eq:totalderivative}. In this case we 
find that $\langle T^\mu_\mu\rangle\ne 0$ 
but very small, of the order of the Hubble parameter, since its value
is controlled by derivatives of the metric. Hence it might appear
that in this case there is no fine tuning problem of the cosmological
constant. However this is not true. In this case the problem arises
in the equation of motion of the scalar field, $\chi$. 
The relevant terms in the equation are,
\begin{equation}
\partial_\mu\partial^\mu\chi + \lambda\chi^3 + ... = 0
\label{eq:scalarfield2}
\end{equation}
If we set the classical value of $\phi=v$, given by 
Eq. \ref{eq:potentialmin}, which minimizes the potential terms involving
the field $\phi$, the remaining
contributions to this equation from the scalar field potential vanish.
We point out that at loop orders the one point function of $\chi$
 will get additional 
contributions. We discuss these below. 
Now the problem is that unless we choose $\lambda$ to be extremely small, 
$\lambda\ll\lambda_2^2$, 
the classical value of $\chi$ will quickly decay to zero. This is true
for a wide range of classical values of $\chi$ that are much larger than the 
electroweak scale. As explained earlier, this is necessarily required
in our framework. 

Let us next determine the order of magnitude of the parameter $\lambda$
that is required in order that $\chi$ varies sufficiently slowly with time
and does not decay to zero over the lifetime of the Universe. We 
assume a space independent scalar field $\chi=\eta$. Hence the time 
scale over which the field decays to zero is given by,
$$ t^2\sim {1\over \lambda\eta^2}$$
If we are to avoid fine tuning of $\lambda$ we should assume that $\lambda$ is
of the order of $\lambda_2^2= v^4/\eta^4$ or larger. 
Furthermore we set $t\sim 1/H_0$
where $H_0$ is the Hubble constant. With this we find that the minimum 
value of $\eta$ is, 
\begin{equation}
\eta \sim {v^2\over H_0}\gg M_{PL}
\label{eq:etasuperPl}
\end{equation} 
Hence if we choose $\eta$ to be of this value, the model will not lead to
a fine tuning problem of $\lambda$.  
This is because all loop contributions to $\lambda$ from the electroweak
sector would be smaller than the value of $\lambda$. This includes the
contributions arising from the vector field kinetic energy terms displayed
in Eq. \ref{eq:actionConformal_d}. 
The fact that $\eta$ is super-Planckian implies that the parameter
$\xi$ in Eq. \ref{eq:SG2} has to be chosen to be extremely tiny.
This might introduce a fine tuning problem in the gravitational 
action. However this action is not renormalizable and the quantum 
theory of gravity not well understood. Hence the standard
measures of naturalness or fine tuning are not really applicable.
In any case we expect that the loop corrections to the gravitational
coupling are very small. 

Let us now consider additional terms which contribute to Eq. 
\ref{eq:scalarfield2}. These arise from the 
conformal regulator in $d$ dimensions. Even if we ignore 
the conformal regulator, there are additional contributions to the one point
function of $\chi$ at loop orders. We must check that these are not 
very large compared with the terms included in Eq. \ref{eq:scalarfield2}.
We first consider the terms arising from the conformal regulator,
taking, as an example, the electroweak gauge boson kinetic energy term 
in Eq. \ref{eq:actionConformal_d}. In $d$ dimensions this introduces a
term in the classical equation of motion of $\chi$ which is proportional to
$$2\delta\, {(\chi^2)^\delta\over \chi}\, \langle \mc E_{\mu\nu} \mc E^{\mu\nu}
\rangle$$
where we have replaced the operator $\mc E_{\mu\nu} \mc E^{\mu\nu}$ 
by its VEV. At the classical level this term is zero. 
The leading order quantum corrections to this expectation value  
are of order $v^4$ and contain a divergence proportional to $1/(d-4)$. 
This divergence is cancelled by the presence of the factor $\delta$ in 
this term. Replacing $\chi$ by its classical value we find that in
the limit $d\rightarrow 4$ this
term gives a contribution of order $v^4/\eta$. This is of the same order 
of magnitude as $\lambda\eta^3$. At higher orders such contributions will
be suppressed by powers of the electroweak coupling. 
Hence such terms give contributions that 
are at most as large as the terms already included. 
By similar analysis we find that the remaining terms from the electroweak
sector also give contributions which are not large compared to the
leading order terms. We consider the remaining loop corrections to 
the one point
function of $\chi$ in section \ref{sec:oneloop}. 

\section{Evading the Fine Tuning Problem} 
\label{sec:solution}
In the previous section we have argued that the fine tuning problem  
is absent
if $\eta$, the classical value of the field $\chi$, is chosen to be
super-Planckian, of order given in Eq. \ref{eq:etasuperPl}. In this section
we argue that a more elegant solution is possible within the framework 
of exact conformal symmetry.  
The solution is based on the presence of a strongly coupled sector which
leads to formation of fermion condensates. These trigger the electroweak
symmetry breaking. For definiteness we consider a specific model 
in which the Higgs is introduced as a fundamental particle. The strongly
coupled sector couples very weakly to the standard model particles 
and hence acts as a dark sector. However other theories, based on 
composite Higgs \cite{Farhi}, might also be considered. It 
should be possible to 
generalize our formalism to such theories also. 

Let us assume the existence of a strongly coupled sector which we
shall refer to as hypercolor. We denote the
gauge fields by $G^a_\mu$ and the corresponding field tensor by 
$G_{\mu\nu}^a$, where $a$ is the internal hypercolor index. The strong gauge
coupling is denoted by $g$. These gauge fields couple to fermions,
$\Psi^i$, where $i$ is the hypercolor index. We assume that these fermions
do not carry electroweak or normal color charge. One can have several
different flavors of these fermions. Here, for simplicity, we assume that
there exists only one flavour of these fermions. 
 The action for this 
strongly coupled sector in d-dimensions may be expressed as,
\begin{equation}
S_S = \int d^dx\sqrt{-g} \left( -{1\over 4} G_{\mu\nu}^a G^{a\mu\nu }
\left(\zeta^2\right)^\delta 
+i\bar\Psi^i \gamma^\mu D_\mu\Psi^i - g_Y\bar\Psi^i\chi\Psi^i\left(\zeta^2\right)^{(-\delta/2)}  \right)
\label{eq:stronglycoupled}
\end{equation}
where $\zeta$ is defined in Eq. \ref{eq:zeta} and
 we have also included a Yukawa coupling of the fermions with 
the scalar field $\chi$. Here the fermion action has to be written
in terms of the vielbein fields due to their coupling with gravity. 
Hence $\gamma^\mu=e^\mu_{\ a}\gamma^a$, where $a$ is a Lorentz index
and $e^a_{\ \mu}$ is a vielbein. 
The model is similar to that introduced
in Refs. \cite{Meissner07,Meissner08,Hur11} within the framework of 
classical conformal 
invariance. 
Now let us assume that, in analogy with QCD, the strongly coupled sector
leads to fermion condensates, i.e. the VEV of the fermion bilinear, 
\begin{equation}
\langle \bar\Psi^i\Psi^i\rangle =\Lambda_S^3 
\end{equation}
where $\Lambda_S$ is the scale of dynamical chiral symmetry breaking. 
As argued in Ref. \cite{Hur11} this will generate a non-zero VEV 
of $\chi$. We can see this from the classical equation of motion for 
$\chi$. This can be expressed as,
\begin{equation} 
\partial_\mu\partial^\mu\chi + \lambda\chi^3 + g_Y\langle\bar\Psi^i\Psi^i\rangle + ... = 0
\label{eq:scalarfield3}
\end{equation} 
Here we have replaced the bilinear $\bar\Psi^i\Psi^i$ by its VEV. 
 In this equation we may, in principle, also have a contribution from the 
gauge fields that arises at loop orders from the gauge kinetic energy term
due to the presence of the $(\zeta^2)^\delta$ regulator. 
These may arise due to the condensate of the gauge fields
\begin{equation}
\langle G_{\mu\nu}^a G^{a\mu\nu}  \rangle =(\Lambda'_S)^4 
\label{eq:hypergluoncondensate}
\end{equation}
where $\Lambda'_S$ is a parameter with mass dimension one.
In order to get an estimate, we separate these contributions
into a non-perturbative and a perturbative part. The non-perturbative
part is represented by the condensate given in Eq. 
\ref{eq:hypergluoncondensate}. This contributes to the equations of
motion in $d$ dimensions due to the $(\zeta^2)^\delta$ term. At the leading
order the contribution vanishes as we let $d\rightarrow 4$. At higher
orders we can only get contributions
 from $\chi$ loops which are highly suppressed. The perturbative
part can lead to additional contributions analogous to the contributions
due to the electroweak gauge bosons. However the gauge bosons of
the strongly coupled sector have zero mass in perturbation theory.
Hence all loop contributions which only involve such gauge particles
with zero external momentum would vanish within the framework of
dimensional regularization and will not contribute. We discuss these
in more detail in section \ref{sec:oneloop}.  

Now the basic point is that the scale of chiral symmetry breaking 
is an independent dimensional parameter. It is governed by the
value of the strong coupling, $g$, at some chosen scale. It is 
approximately equal to the energy scale at which $g$ becomes larger
than unity. We clarify that despite quantum conformal invariance, the
coupling parameters do depend on scale in this framework 
\cite{Shaposhnikov:2008a,Tamarit13}.  
In the $d$ dimensional action the strong hypercolor gauge field 
kinetic energy terms are coupled to the scalar field. As discussed
earlier, we treat these terms by expanding the
field $\chi$ about its classical value $\eta$. So far, however, the
field $\eta$ is undetermined. It serves the same purpose as the
scale $\mu$ that is normally introduced in dimensional regularization
in order to account for the mass dimension of the couplings, that are 
dimensionless in $d=4$ but become dimensional when $d\ne 4$. 
The strong coupling dynamics does not really relate the scale $\mu$,
or equivalently $\eta$, to the scale of chiral symmetry breaking. 
In order to see this more explicitly, consider the running strong
 coupling
parameter. In the perturbative regime, its value at some chosen scale
$\mu_0$ has to be specified. The renormalization group equation
then gives its value at the scale $\mu$. 
  Here the scale $\mu$ which appears
in the $d$ dimensional action is completely arbitrary. The scale $\mu_0$
is some choice made at which the value of $g$ is being measured. 
The order of magnitude of the 
parameter $\Lambda_S$ is equal to the mass scale at which the 
strong hypercolor coupling
becomes larger than unity. It is clear that this is not directly related
to $\mu$ (or $\eta$ is our case) as long as we only consider the strong
interaction dynamics. However $\eta$ gets related to $\Lambda_S$ by
the equation of motion of the scalar field, Eq. \ref{eq:scalarfield3}.  
Hence,  within the framework of the conformal
regularization, i.e. GR-SI prescription, the arbirary scale introduced by 
the regularization procedure gets related to the scale of strong
interaction dynamics by Eq. \ref{eq:scalarfield3}. 

Our proposal solves the fine tuning problem of $\lambda$ since now 
its value need not be very small.
We can choose it to be of order $\lambda_2^2$ or even larger.
Furthermore we do not require that
$\eta$ is super-Planckian. At leading order Eq. \ref{eq:scalarfield3}
implies that the solution is
\begin{equation}
\eta^3 = -{g_Y\over \lambda} \langle\bar\Psi^i\Psi^i\rangle
\end{equation} 
where $\eta$ is constant. 
This is similar to the generation of the vacuum expectation value of the
scalar field in Ref. \cite{Hur11}. The main difference in our case is that the
scalar field itself is being used as a dynamical regulator.
At higher order we will get additional contributions to the one point
function of $\chi$ which can shift the classical value of $\chi$. However
the important point is that we can obtain a consistent solution with $\eta$ 
equal to a constant at each order in perturbation theory without requiring
any fine tuning. The main problem of fine tuning lies in the smallness 
of the value of $\lambda$. Now this parameter is taken to be sufficiently
large so that it does not require fine tuning.  
Once the field $\chi$ acquires non-zero classical value, 
the electroweak symmetry is also broken with $v$ related
to $\eta$ by Eq. \ref{eq:potentialmin}. The scale $\Lambda_S$ is related
to the electroweak scale only through the field $\chi$. This strongly 
coupled sector acts as a dark matter sector since its couplings with
visible matter arise only through $\chi$ and are highly suppressed.

We emphasize that there do exist two very small parameters in the model. 
These are $\lambda_2= v^2/\eta^2$ and $\lambda\sim \lambda_2^2$ or $\lambda
>\lambda_2^2$. 
Essentially $\lambda_2$ is small since we need to choose
$\sqrt{\lambda_2}=v/\eta\ll 1$. This parameter controls the coupling 
of the visible sector with the dark matter sector. 
Strictly speaking, the parameter $\lambda$ need not be very tiny. 
It can take any value larger than $\lambda_2^2$. We shall later
discuss the implications of different choices of this parameter. 

Based on the gravitational action,
Eq. \ref{eq:SG1},
it has earlier been argued that $\eta$ should
be of the order of $M_{PL}$ \cite{Cheng88,JMS,Shaposhnikov:2008a,Shaposhnikov:2008b}. However as we have discussed above this is not
well motivated. Another motivation for this choice is to suppress 
the loop contributions 
due to the regulator terms, such as, $(\zeta^2)^\delta$ in the
gauge field kinetic energy terms. These contributions are expected
to be non-renormalizable at two loops and we would prefer the
scale of non-renormalizability to be very high, perhaps of the
order of Planck mass. However so far there does not exist an explicit
proof of this absence of renormalizability. We will not address this
issue in the present paper and simply assume that $v/\eta\ll1$.  
 
We can get an estimate of the strong hypercolor coupling scale, $\Lambda_S$, by
setting $\lambda\sim v^4/\eta^4$, which is roughly the minimum value
it can take in order to avoid fine tuning. We obtain, 
\begin{equation}
\Lambda_S \sim v \left({v\over g_Y\eta}\right)^{1/3}
\label{eq:estimateLS}
\end{equation}
Assuming that $g_Y$ is somewhat smaller than unity, but not very tiny, 
we find that $\Lambda_S$ is a few orders of magnitude smaller than $v$. 
For $\lambda$ of order unity, we find that $\Lambda_S\sim \eta/g_Y^{1/3}$.

We should point out that in the presence of several flavors of the 
strongly coupled fermions, Eq. \ref{eq:scalarfield3} will get 
contributions from all the flavors. Since these may involve different
Yukawa couplings, our estimate of $\Lambda_S$ in Eq. \ref{eq:estimateLS} 
may differ significantly in this case. The one point function of the
field $\chi$ also gets corrections at loop orders from the QCD and the
electroweak sector. However since $\lambda$ is sufficiently large,
this do not cause any fine tuning problems. 
The one loop contributions are discussed in the next section.

\section{One loop contributions}
\label{sec:oneloop}
In this section we explicitly demonstrate that the one loop corrections
do not lead to any fine tuning of parameters. As described earlier 
the main issue is the presence of one or two very small parameters in the
theory. These are $\lambda_2\sim v^2/\eta^2$ and $\lambda>\lambda_2^2$. 
As we have explained earlier, $\lambda$ is not necessarily very small.
It can take any value larger than $\lambda_2^2$. 
As we shall see the loop corrections to these parameters are small.
Here we shall be primarily interested in the contributions due to the
scalar field potential, the electroweak sector and the
strong coupling hypercolor sector.  
 We emphasize that
the dark matter sector couples to electroweak particles only through
the field $\chi$. This coupling is very weak. In fact it is precisely this
weak coupling whose stability we are interested in testing. 
Furthermore the field $\chi$ couples to the dark matter sector through
the Yukawa couplings. Hence we expect some constraints on these couplings
so that these do not 
 destabilize the scalar field potential.

We first carry out our one loop analysis assuming the extreme case of 
$\lambda\sim\lambda_2^2$. In this case $\lambda$ is very very small. 
We shall consider the other limit of $\lambda\gg\lambda_2^2 $ later.
We expand the fields $\chi$ and $\phi$ around
their classical values, $\eta$ and $v$ respectively, 
\begin{eqnarray}
\chi &=& \eta + \hat\chi\nonumber\\
\phi &=& v + \hat\phi
\end{eqnarray}
After substituting this in the potential, we find that the fields $\hat\phi$
and $\hat\chi$ undergo a small mixing. At leading order, 
the mixing parameter is given by,
\begin{equation}
\sin\theta \approx \sqrt{\lambda_2} 
\end{equation}
The physical fields are identified as
\begin{eqnarray}
\tilde\chi &\approx& \hat\chi + 
\sqrt{\lambda_2}\hat\phi \nonumber\\
\tilde\phi &\approx&  \hat\phi - 
\sqrt{\lambda_2}  \hat\chi
\label{eq:tildefields}
\end{eqnarray}
The particles $\tilde\phi$ and $\tilde\chi$ have masses given by,
\begin{eqnarray}
m^2_\phi &\approx& 2\lambda_1 v^2\nonumber\\
m^2_\chi &\approx& 3\lambda\eta^2 
\end{eqnarray}
We clarify that we expect one massless scalar field due to soft breaking
of conformal invariance. We expect this to be dominantly 
a mixture of the field $\tilde
\chi$ and a bound state of the dark strongly coupled fermions. Here we have 
ignored this sector with the assumption that the 
mixing is small due to the smallness of the Yukawa couplings. 
In any case we find that the mass of $\tilde\chi$ is very very small,
$m_\chi\ll m_\phi$. Here we are only interested in order of magnitude
estimates to ensure that the one loop corrections do not 
lead to fine tuning of parameters. For this purpose we are justified
in ignoring this small mixing. 

In terms of these fields, the relevant terms which arise in the Lagrangian
due to the scalar potential are, 
\begin{equation}
{\cal L} = -\lambda_1v\tilde\phi^3 - {\lambda_1\over 4}\tilde\phi^4
-2\lambda_1\lambda_2\eta\tilde\phi^2 \tilde\chi- \lambda\eta \tilde\chi^3
- {\lambda\over 4}\tilde\chi^4 - \lambda_1\lambda_2\tilde\phi^2\tilde\chi^2
-\lambda_1\sqrt{\lambda_2}\, \tilde\phi^3 \tilde\chi + ... 
\end{equation}
To maintain conformal invariance in $d$ dimensions, we need to modify 
the potential as,
\begin{equation}
\mc L_V =  -\left[{\lambda\over 4}\chi^4+{\lambda_1\over 4}\left(\phi^2-\lambda_2\chi^2\right)^2\right] \left(\zeta^2\right)^{-\dl}
\label{eq:Vchiphiddim}
\end{equation}
The Feynman rules corresponding to this Lagrangian are given in the Appendix.
There we have also shown the rules for the terms arising 
from the electroweak gauge field kinetic energy term in $d$ dimensions,
\begin{equation}
\mc L_{EW} = -\frac14\left(F_{\m\n}^a F^{a\,\m\n}\right)\left(\zeta^2\right)^\dl
+ ...
\label{eq:LEWdm}
\end{equation}
where $F_{\m\n}^a$ represents the electroweak gauge field tensor. 
We also include the rules for the Yukawa interaction terms 
 in $d$ dimensions. 
We might expect significant contribution from the top quark Yukawa interaction.
Here we assume a heavy fermion coupled to the scalar field $\phi$, which
is representative of the Higgs field in our model. The analysis may
also be carried out within the framework of the standard model but
this is expected to be similar to the analysis in our toy model. 
The Lagrangian for the Yukawa term may be written as,
\begin{equation}
\mc L_Y = 
- g_t\bar t\phi t\left(\zeta^2\right)^{(-\delta/2)}  
\label{eq:yukawatt}
\end{equation}
where $t$ represents a heavy fermion of mass equal to the mass of the top quark.

\begin{figure}[h]
\begin{center}
\scalebox{0.5}{\includegraphics*[angle=0,width=\textwidth,clip]{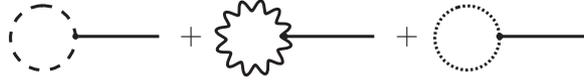}}\end{center}
\caption{The diagrams contributing to the one point function of $\tilde\chi$. 
From left to right the diagrams involve a $\tilde\phi$ loop, an electroweak
gauge particle loop and a top quark loop.
} 
\label{fig:chionepoint}
\end{figure}
 
\begin{figure}[h]
\begin{center}
\scalebox{0.5}{\includegraphics*[angle=0,width=\textwidth,clip]{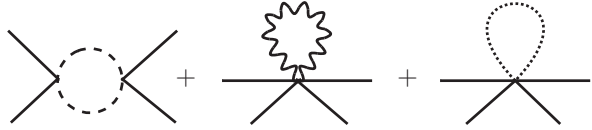}}\end{center}
\caption{The diagrams contributing to $\tilde\chi
\tilde\chi\rightarrow \tilde\chi\tilde\chi$ scattering. From left to
right the diagrams
involve a $\tilde \phi$ loop, an electroweak gauge boson loop and a top loop.
} 
\label{fig:chichiscattering}
\end{figure}

\begin{figure}[h]
\begin{center}
\scalebox{0.5}{\includegraphics*[angle=0,width=\textwidth,clip]{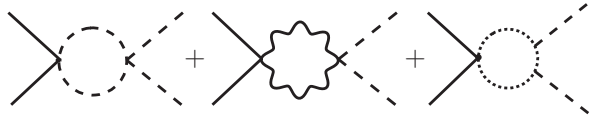}}\end{center}
\caption{The diagrams contributing to $\tilde\chi
\tilde\chi\rightarrow \tilde\phi\tilde\phi$ scattering. Here the left
diagram involves a $\tilde\phi$ loop. The middle involves an electroweak 
gauge boson loop which go into a pair of Higgs bosons, which are represented 
here by the $\phi$ particles, through the
standard electroweak vertex. The right diagram involves a top quark loop.
The top quark emits two Higgs particles due to the Yukawa interaction.  
} 
\label{fig:chiphiscattering}
\end{figure}

We now check the order of magnitude of the one loop contributions. As 
explained earlier we are basically interested in determining the order 
of magnitude of the finite parts. If these are too large then we will
require large counter terms which will lead to fine tuning. Let us first 
consider the one point function of the field $\tilde \chi$. The one
loop corrections to this are shown in Fig. \ref{fig:chionepoint}  
The first term arises directly from the potential and is of order
$\lambda_1\lambda_2 v^2\eta$. This may be compared with the 
classical contribution
to this amplitude which arises from the $\lambda\chi^4$ term. 
The relevant term is proportional to $\lambda\tilde\chi\eta^3$ and 
gives a contribution of order $v^2\eta\lambda_2$. 
It is clear that the loop correction is smaller than the leading order term.
We also get contributions to the one point function due to the
$\mc L_{EW}$ and $\mc L_Y$ Lagrangians displayed above. The one loop 
contributions arise due to the diagrams shown in Fig. \ref{fig:chionepoint}. These 
contributions are found to be finite due to the presence of $\delta$ in
the relevant Feynman rule in $d$ dimensions. The contributions of
these terms are found to be of order $v^4/\eta\sim v^2\eta\lambda_2$
and hence are found to be the same order as the leading order term. 
Here we have set $M_W\sim m_t\sim v$, where $M_W$ and $m_t$ are the
W-boson and top quark masses respectively. The higher order contributions
of these terms will be suppressed by additional power of weak coupling or
$g_t$. Hence we see that these do not lead to any fine tuning. 

An important point is that 
there is no mass term in the Lagrangian due to conformal 
invariance. Hence the scalar field mass is not an independent parameter
and we do not need to check the corrections to this parameter. In 
particular the Higgs mass is 
expected to be stable under quantum corrections due to conformal invariance 
\cite{Shaposhnikov:2008a,Tavares14}.

We next look at the four point function corresponding to $\tilde \chi$. 
The relevant one-loop corrections are shown in Fig. \ref{fig:chichiscattering}.
 The first term is
of order $(\lambda_1\lambda_2)^2$ which is small compared to $\lambda$. 
The other two terms are of order $v^4/\eta^4$ which is of the same 
order as $\lambda$. Hence none of these lead to any fine tuning problems.
As in the case of the one point function, higher order contributions 
from all the terms would be further suppressed.
This clearly shows that the parameter $\lambda$ does not require any
fine tuning. 

We next check the parameter $\lambda_2$. This contributes to the scattering
process $\tilde \chi\tilde\chi\rightarrow\tilde \phi\tilde\phi$. 
The relevant one loop diagrams are shown in Fig. \ref{fig:chiphiscattering}. 
The first diagram
gets contribution directly from the potential and is of order $\lambda_1^2
\lambda_2$ and hence suppressed compared to the leading order amplitude. 
The remaining diagrams arise due to the presence of the conformal regulator.
As in the case of $\tilde \chi\tilde \chi\rightarrow \tilde \chi\tilde \chi$,
these diagrams give finite contributions, which are smaller or comparable
to the leading order amplitude. Furthermore at higher orders such contributions
will be suppressed compared to the leading order result. Hence we see
that the model does not require any fine tuning of parameters.  

\begin{figure}[h]
\begin{center}
 \scalebox{0.4}{\includegraphics*[angle=0,width=\textwidth,clip]{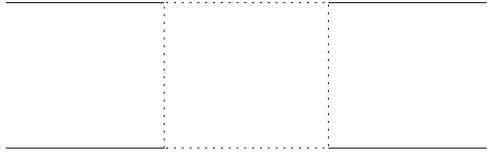}} 
\end{center}
\caption{The box diagram contribution to the $\tilde\chi\tilde\chi\rightarrow 
 \tilde\chi\tilde\chi$
scattering. The internal lines correspond to fermions in the dark strongly 
coupled sector.} 
\label{fig:chi4}
\end{figure}

We next check the contributions due to the hidden strongly coupled sector. 
The $\chi$ field also couples to the hidden fermions by the Yukawa 
interaction terms. These terms can contribute to the 
$\tilde\chi\tilde\chi\rightarrow \tilde\chi\tilde\chi$ by the diagram shown in Fig. \ref{fig:chi4}.
This diagram gives a contribution of order $g_Y^4$. Hence we should require
that $g_Y<\lambda^{1/4}\sim \lambda_2^{1/2}\sim v/\eta$. We find that
in this limit, the minimum value of $\Lambda_S$ is given by,
 $\Lambda_S\sim v$. 
Such a small value of Yukawa couplings have interesting implications
for the hidden sector meson spectrum. We point out that this sector 
has approximate chiral symmetry which is broken by the Yukawa couplings. 
In the limit of exact symmetry we expect zero mass pseudoscalar Goldstone bosons
corresponding to broken global chiral symmetry. However due to explicit
breaking these will acquire small masses. The interesting point
is that the breaking a very tiny and hence the mass is expected
to very very small in comparison to the scale $\Lambda_S\sim v$. 
This implies existence of dark matter particles whose masses are much 
smaller than the electroweak scale.

We next consider the contributions from the hidden gauge sector.  
As we have shown above, 
for our typical choice of parameters $\lambda\sim\lambda_2^2$, $\lambda_2
\sim v/\eta$, the minimum value of $\Lambda_S\sim v$.  
We expect that the mass scale of the gauge sector, $\Lambda'_S$, to be
also of this order. Hence all loop contributions would at most be of
the same order as those obtained from the electroweak sector. As we
have discussed in section \ref{sec:solution}, in order to evaluate
these contributions we should split them into a non-perturbative and
a perturbative part. Furthermore we have argued that
the non-perturbative contributions are suppressed. Treating these 
gauge particles perturbatively we get contributions analogous to those
obtained from the electroweak gauge bosons, as shown in Figs. 
\ref{fig:chionepoint}, \ref{fig:chichiscattering}, \ref{fig:chiphiscattering}. In order to test
whether these lead to any fine tuning, we can set the external momentum 
equal to zero. Furthermore these gauge particles are massless within
the framework of perturbation theory. Hence all the loop contributions 
analogous to those shown in  Figs. 
\ref{fig:chionepoint}, \ref{fig:chichiscattering}, \ref{fig:chiphiscattering} vanish within 
dimensional regularization. This is true also for all loops which involve
only these particles. Loops which involve these particles along with
other particles can lead to non-zero contributions. The dominant
contributions are expected to arise from loops involving the hidden
gauge particles and hidden fermions. The perturbative masses of
these hidden fermions are of order $g_Y\eta<v$. Hence we expect these
masses to be small compared to $v$. In this case these loops cannot give
contributions larger in comparison to those obtained from the electroweak
sector. As argued above these are sufficiently small and do not lead to
any fine tuning problems. 

We next consider the limit $\lambda\gg\lambda_2^2$. We assume that $\lambda$
is small compared to unity but sufficiently large so that there is
no issue of acute fine tuning of this parameter. 
The box diagram shown in Fig. \ref{fig:chi4} leads to an amplitude
of order $g_Y^4$. If $g_Y$ is of order unity and $\lambda$ very small compared
to unity, this might require a mild fine tuning at one loop order. In 
order to avoid this we might demand that $g_Y<\lambda^{1/4}$.  
In this case we find that $\Lambda_S> g_Y\eta$. Assuming that $g_Y\gg 
\sqrt{\lambda_2}$, this implies that $\Lambda_S\gg v$. Hence in this case
the mass scale of the hidden strong sector is much larger than the 
electroweak mass scale. Since $\lambda\gg\lambda_2^2$, the mixing
of $\tilde \chi$ with $\tilde\phi$ is negligible. We find that the
mass of $\tilde\phi$ (or Higgs) particle is given by $m_\phi^2
\approx 2\lambda_1v^2$ and mass of  $\tilde\chi$, $m_\chi^2\sim
\lambda\eta^2$. Hence we find that in this case, $m_\chi\gg m_\phi$.
A precise calculation of $m_\chi$ is complicated since 
$\tilde\chi$ also mixes with a scalar bound state of the strongly
coupled hidden sector. Due to soft breaking of the conformal symmetry
we expect one scalar particle of zero mass and another of order, 
$m^2\sim \lambda\eta^2$. The only very small parameter in this model is
$\lambda_2\sim v^2/\eta^2$. This parameter essentially couples the
field $\chi$ with the electroweak sector. At one loop the diagrams which
can contribute to this parameter are shown in Fig. \ref{fig:chiphiscattering}.  
It is easy to see that none of them lead to a large correction which
may require fine tuning of $\lambda_2$.

Before ending this section we briefly consider the one loop corrections
to the one point function of $\tilde\chi$ within the framework of the
model with a super-Planckian value of $\eta$ discussed in section 
\ref{sec:gravity}. The relevant diagrams are shown in Fig.  
\ref{fig:chionepoint}. We find that the maximum value of these contributions
is of order $v^4/\eta$, which are comparable to the leading order
contributions. At higher order these will be further suppressed
by powers of the electroweak coupling or the $\phi$ (or Higgs) self coupling
or the top Yukawa coupling. Hence these contributions do not lead to 
any fine tuning of the parameter $\lambda$ in this model.  

\section{A constraint on the scale $\eta$}
\label{sec:constraint}
\begin{figure}[h]
\begin{center}
 \scalebox{0.5}{\includegraphics*[angle=0,width=\textwidth,clip]{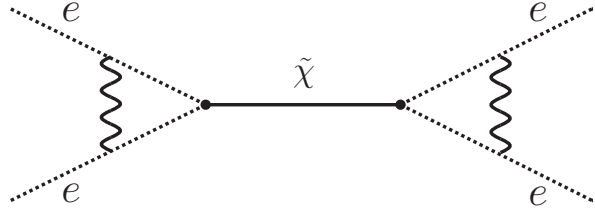}} 
\end{center}
\caption{The electron-electron scattering by exchange of a scalar field $\tilde \chi$. The dotted lines represent an electron. The wavy lines represent
a photon or a weak gauge boson. 
 Here we show only a $s$-channel diagram. 
We can also have a $t$ and a $u$-channel process. 
} 
\label{fig:econstraint}
\end{figure}

In this section we point out a constraint on the parameter space which
arises due to the presence of the conformal regulator. This has so far 
not been realized in the literature. Let us consider the standard model
Yukawa interaction term of the electron. 
This can be expressed as,
\begin{equation}
\mc L_Y = -\left[g_e (\bar \nu_e\  \bar e)_L {\mc H} e_R + h.c. \right]\left(\zeta^2\right)^{-\delta/2}
\end{equation}
Here $\mc H$ denotes the Higgs field and $\nu_e$ and $e$ the neutrino and
electron fields respectively.  
Here we use the standard model Higgs field in contrast to the toy singlet
Higgs, $\phi$, used in the rest of this paper. 
We need to suitable modify the form of $\zeta$ in Eq. \ref{eq:zeta}
by replacing $\phi^2$ by ${\mc H}^\dagger\mc H$. 
We expand $\mc H$ about its vacuum expectation value $v$. 
 This leads to the mass term for the electron. 
Now the important point is that in $d$ dimensions these mass terms
couple to the field $\chi$. This coupling leads to a non-zero contribution
to the scattering amplitude $ee\rightarrow ee$ displayed in Fig. 
\ref{fig:econstraint}. 
The wavy lines represent a photon or a weak gauge boson. 
We consider loops instead of the tree level process because the $ee\chi$
coupling is proportional to $(d-4)$ and hence goes to zero when $d\rightarrow
0$. However, with the two loops, the $(d-4)$ factors get cancelled by the
$1/(d-4)$ factors arising due loop divergences. 
We are interested
in scattering at low energy for which the dominant contribution arises 
from the exchange of photons. 

The contribution from each loop in $d$ dimensions is proportional to,
$$\int d^dp\, {1\over p^2}\, {p^2\over (p^2-m_e^2)^2}$$
where $m_e$ is the mass of the electron.
In $d=4$ this integral is dimensionless. 
 Here we have set the external 
momentum equal to zero. Each of the two $ee\chi$ vertices yield a factor
of $m_e/\eta$. Besides that each photon exchange yields a factor of
$\alpha$. Hence the order of magnitude of this amplitude is 
$(\alpha m_e/\eta)^2$. This is small but not entirely negligible since 
the coupling involves the mass of the electron and
the effect accummulates in the case of macroscopic bodies.  

\begin{figure}[h]
\begin{center}
 \scalebox{0.5}{\includegraphics*[angle=0,width=\textwidth,clip]{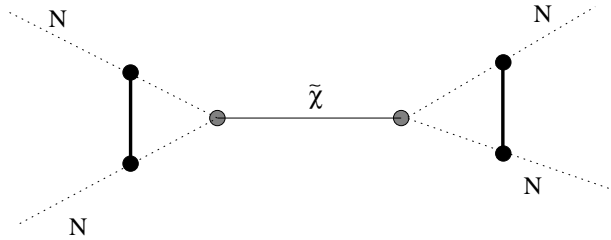}} 
\end{center}
\caption{The nucleon-nucleon scattering by exchange of a scalar field $\tilde \chi$. The dotted lines represent a nucleon. The thick solid lines represent
a multigluon exchange or a meson such as a pion or a $\rho$ meson etc. 
These may also represent exchange of a photon or a weak gauge boson.
 The blobs represent
effective vertices which involve the nucleon form factors. 
} 
\label{fig:constraint}
\end{figure}

Before discussing the implications of such amplitudes we consider 
the corresponding diagrams for the case of hadrons. In principle, 
such contributions may also arise for nucleons. A possible diagram
is displayed in Fig. \ref{fig:constraint}. 
The $NN\tilde\chi$ coupling arises due to the Yukawa 
interaction terms of the up and down quarks. Here we require an 
effective coupling of the nucleon with $\chi$ which will involve a
form factor. 
The thick solid lines in Fig. \ref{fig:constraint} represent exchange of
strongly interacting particles or a photon or a weak gauge bosons. 
For simplicity let us first consider a photon. We again set the external 
momentum equal to zero. At each of the photon vertices we need to 
include the nucleon electromagnetic form factor which decays as $1/p^4$ 
for large values of $p^2$. Inserting such a form factor into the loop 
integral, we find that the integral is finite and hence the total 
contribution from such diagrams vanish in the limit $d\rightarrow 4$.    
We expect the presence of similar form factors for the case of
multi-gluon or meson exchanges also and hence such amplitudes vanish 
in the case of nucleons.

For macroscopic bodies we expect a small deviation from the
standard law of gravity arising due to the coupling of $\chi$ to the mass
of electrons. This additional contribution is very tiny if we choose 
$\eta\sim M_{PL}$. In this case the amplitude and hence the corresponding 
potential will be suppressed by more than 10 orders of magnitude 
in comparison to the standard gravitational potential. This is due to 
factors of $\alpha^2$ and an additional suppression factors of
$m_e/m_N$ in comparison to standard gravitational interaction. Here
$m_N$ is the mass of the nucleon.  
Furthermore the particle $\tilde \chi$ is massive with mass equal to 
$3\lambda\eta^2$. With $\lambda\sim v^4/\eta^4$ and $\eta\sim M_{PL}$, 
this leads to a mass of about $10^{-6}$ eV. Hence the interaction 
is appreciable only over a distance of a few cm. It is clear that 
this choice of parameters are allowed. However if $\eta$ is taken to
be much smaller than it might conflict with experimental constraints. 
We postpone a detailed discussion of such contraints to future work.

We point out that there exists a massless
scalar particle in our theory due to soft breaking of scale invariance.
This particle is expected to be a mixture of $\tilde \chi$, $\tilde\phi$
and a bound state of the dark sector fermions. It can also contribute 
to the amplitudes shown in Figs. \ref{fig:econstraint} and 
\ref{fig:constraint} due to its mixing with $\tilde \chi$. 
In this case the interaction would have an infinite range since the
particle being exchanged is massless. However we expect that the mixing
is small and hence the amplitude would be further suppressed in comparison
to the amplitude due to the exchange of $\tilde \chi$. We also point
out that such amplitudes would be negligible if we impose local 
conformal invariance since in this case the massless particle would
not be present in the physical spectrum and act as the longitudinal
mode of the Weyl meson. Furthermore all our arguments with regards
to the absence of fine tuning of parameters $\lambda$ and $\lambda_2$ 
apply even in the case of local conformal invariance. 

\section{Including the Standard Model fields}
In our discussion so far we have considered a toy model in which the
Higgs multiplet is represented by a real scalar field. However 
the main results of our paper are applicable in the full standard model
also. We briefly illustrate this in this section.
The conformal extension of the standard model in $d$ dimensions can be
expressed as, \cite{JM10_2}. 
\begin{equation}
{\cal S} = 
 {\mc S}_G  + {\mc S}_{C} 
\label{eq:LagrangianSM}
\end{equation}
where ${\mc S}_G$ is the gravitational action and ${\mc S}_{C}$ 
represents the conformal extension of the standard model. It is given by,
\ba
\mathcal{S}_{SM} &=& \int d^dx \sqrt{-g}\Bigg[{1\over 2}g^{\mu\nu}\partial_\mu\chi
\partial_\nu\chi+ 
 g^{\mu\nu} (D_\mu \mc H)^\dag(D_\nu \mc H) -\frac14
g^{\mu\nu} g^{\alpha\beta}(\mathcal{A}^i_{\mu\alpha} \mathcal{A}^i_{\nu\beta}
\nn\\
&+& \mathcal{B}_{\mu\alpha} \mathcal{B}_{\nu\beta} + \mathcal{G}^j_{\mu\alpha}
 \mathcal{G}^j_{\nu\beta}) (\zeta^2)^{\delta}
 - {\lambda_1\over 4}   (4\mc H^\dag \mc H-\lambda_2\chi^2)^2 
(\zeta^2)^{-\delta}-{\lambda\over 4}\chi^4(\zeta^2)^{-\delta}  \Bigg]\nonumber\\
 &+& {\cal S}_{\rm fermions},
\label{eq:S_EW_d}
\ea
where $\mc H$ is the Higgs multiplet and $\mc{G}^j_{\mu\nu}$, $\mc{A}^i_{\mu\nu}$ and
$\mc{B}_{\mu\nu}$ and $\mc{E}_{\mu\nu}$
are the standard field strength tensors for the $SU(3)$, $SU(2)$ and $U(1)$
gauge fields. 
Here $\zeta$ is same as given in Eq. \ref{eq:zeta} with $\phi^2 $ replaced
by ${\mc H}^\dagger\mc H$. 
We point out that these gauge fields remain unchanged
under conformal transformation.  
The fermion action, ${\cal S}_{\rm fermions}$, is given by
\ba
{\cal S}_{\rm fermions} &=&\int d^d x\, e\, \left({\overline\Psi}_{\rm L}i
\gamma^\mu  {\cal D_\mu} \Psi_{\rm L} +
{\overline\Psi}_{\rm R}i \gamma^\mu  {\cal D_\mu} \Psi_{\rm R}  \right)\nn\\
&&- \int d^dx\, e\, (g_Y\overline{\Psi}_{\rm L} {\mc H}\Psi_{\rm R} 
(\zeta^2)^{- \delta/2} + h.c.),
\label{eq:Sfermions}
\ea
where $e={\rm det}(e_\mu^{~a})$, $e_\mu^{~a}e_\nu^{~b}\eta_{ab}
=g_{\mu\nu}$, $\gamma^\mu=e^\mu_{~a}\gamma^a$, $e_\mu^a$ is the
vielbein  and
$a, b$ are Lorentz indices. 
Here $\Psi_L$ and $\Psi_R$ represent the left and right handed projections
of the fermion field and $g_Y$ is a Yukawa coupling. The gauge
covariant derivatives
of fermion fields are same as in the standard model. 
Here also we have assumed global conformal invariance but the model can
be easily generalized to display local invariance.
In Eq. \ref{eq:Sfermions} we have shown only a representative term of the
fermion field. Additional terms corresponding to different Yukawa couplings
and different families can be added analogously.

It is clear from the action that the analysis presented in sections
\ref{sec:problem} and \ref{sec:solution} 
is applicable in this case also. Here we have essentially replaced the
real scalar $\phi$ with the Higgs field. The Higgs field acquire VEV 
 once the classical value of $\chi$ is non-zero. 
As discussed in section \ref{sec:problem}, we can evade the fine tuning
problem if we choose the classical value of $\chi$ to be much
larger than the Planck mass. Alternatively, as discussed in section
\ref{sec:solution} we add a dark strongly coupled sector, which triggers
the breakdown of electroweak symmetry through the vacuum expectation
value of $\chi$.

\section{Conclusions}
In this paper we have explained how the fine tuning problem of the 
cosmological constant manifests itself in a conformal model. 
We have assumed a matter action which displays global conformal invariance
in $d$ dimensions. In such a model the trace of the energy momentum
tensor is zero even for the dimensionally regulated action. Hence
the model leads to vanishing cosmological constant even at loop orders.
However the problem of fine tuning of cosmological constant manifests
itself in the requirement to break the conformal symmetry by a soft 
mechanism. In the simplest version of the model, this 
is implemented by generating a non-zero classical value of the real scalar field
$\chi$, which in turn leads to a breakdown of the electroweak symmetry.
 This mechanism requires two 
very small parameters. One of these parameters, denoted by $\lambda_2$ in 
our paper, is equal to $(v/\eta)^2$, where $\eta$ is some very large
mass scale, which may be taken to be $M_{PL}$. Although this parameter
is small, we find that it does not get large corrections at loop orders
and hence does not require fine tuning. However there exists another
parameter, denoted as $\lambda$ in our paper, which leads to acute
fine tuning problems. If this parameter is not fine tuned, the 
field $\chi$ quickly decays to zero and the conformal symmetry is restored.
We argue that this situation is avoided if the classical value
of $\chi$ is much larger than $M_{PL}$.  

We also discuss another model in which there exists 
a dark strongly coupled sector. In this case the conformal symmetry 
breaking is triggered by the formation of condensates in this sector.
In this case also the full model, including the hidden as well as the
standard model sector displays exact conformal invariance in $d$ dimensions. 
The generation of condensates lead to a non-zero vacuum value of $\chi$
which leads to a breakdown of electroweak symmetry. With this mechanism
also we find that none of the parameter suffer from fine tuning problems
at loop orders. 
We describe the
parameter ranges over which this is applicable. 
We have discussed a specific model which involves a strongly coupled 
sector. However our mechanism may be applicable to other models also
which involve interactions similar to technicolor provided the
conformal invariance is maintained within the regulated action.
Furthermore it may also be interesting to consider the model 
with local conformal invariance.  
The mechanism discussed
in our paper should be applicable in this case also. 

In our analysis we have focussed primarily on the problem associated with
the fine tuning of the cosmological constant. We have shown that it
is possible to set it to zero without having to arbitrarily set some
parameter to zero order by order in perturbation theory. We have so 
far not addressed the issue of how the dark energy and dark matter may
be generated in our model. Our main point is that the class of models
which we present provide us with a useful starting point in which to 
address these issues.  
There does exist a dark sector which might
be responsible for these components.   
However the problem is a little complicated since the trace of the
energy momentum tensor and hence $R$ is very close to zero in these models.
The deviation from zero is controlled by the gravitational action which
is not conformally invariant or is obtained by making a particular gauge
choice within the framework of local conformal invariance \cite{Padmanabhan85}.
It is clear that such terms will generate a value of $R$ of the order
that is acceptable by cosmological considerations. However more work
is required in order to fit the cosmological data within this framework.

\section{Appendix: Feynman Rules}

In this Appendix we summarize the Feynman rules in our theory. These 
correspond to the Lagrangian densities, $\mc L_{EW}$, $\mc L_V$ and 
$\mc L_Y$ displayed in Eq. \ref{eq:LEWdm}, 
\ref{eq:Vchiphiddim} and \ref{eq:yukawatt} respectively. 
The field $\zeta$ is defined in Eq. \ref{eq:zeta} and we need to expand 
terms such as, $(\zeta^2)^\delta$. The leading order terms in such an
expansion are displayed in Eq. \ref{eq:expepsi}. 
Here the dominant contributions arise from the $\hat\chi$ terms and
hence we ignore the $\hat \phi$ terms. We next express $\hat \chi$
in terms of $\tilde \chi$ using Eq. \ref{eq:tildefields}. 
At leading order the two are equal to one another. 

We denote the $\tilde \chi$ and $\tilde\phi$ lines by a solid and dashed
lines respectively. An electroweak gauge boson is denoted by a wavy line
and a fermion by a dotted line as shown in Fig. \ref{fig:notation}. 
The Feynman rules for the couplings of $\tilde\chi$ which arise from the Lagrangian given in Eq. \ref{eq:LEWdm}
are shown in Figs. \ref{fig:EW1}, \ref{fig:EW2}, \ref{fig:EW3}.  
The corresponding rule due to the Yukawa term, Eq. \ref{eq:yukawatt}, is given
in Fig. \ref{fig:Yukawa1}. 
Finally the Feynman rules arising from the scalar field potential terms
are given in Fig. \ref{fig:potential}.

\begin{minipage}{\textwidth}
\centering
\includegraphics*[angle=0,width=0.4\textwidth,clip]{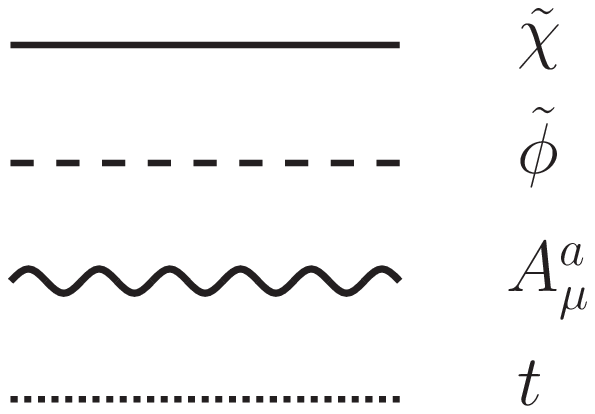}
\end{minipage}
\captionof{figure}{
The conventions used in the Feynman diagrams. Here $A_\mu^a$ represents
an electroweak gauge boson.
}
\label{fig:notation}
%\section*{Feynman rules from EW interaction}
%\begin{figure}[h]
\begin{minipage}{0.4\textwidth}
\centering
 \includegraphics[width=0.7\linewidth]{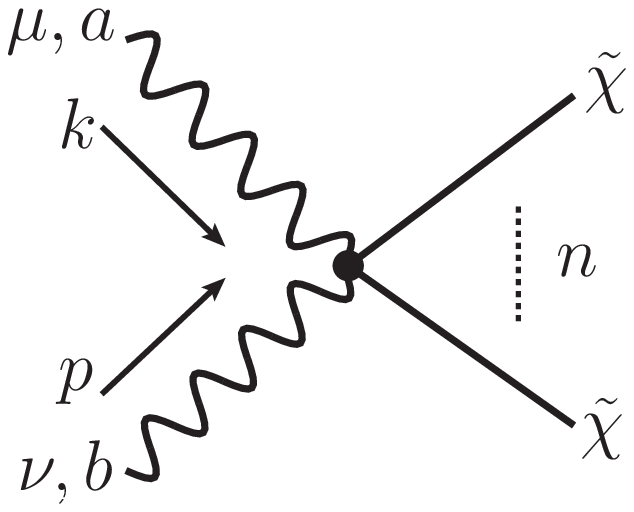} 
 \end{minipage}\hfill
\begin{minipage}{0.6\textwidth}
\centering
\begin{equation*}
 =-i\dl(-1)^n \frac{2 (n-1)!}{\et^{(n-2\dl)}}\times\Big(2\left(k\cdot p\right)g^{\m\n} - k^\m p^\n - k^\n p^\m\Big)
\end{equation*}
 \end{minipage}\\
 \captionof{figure}{The scattering of two electroweak gauge bosons with $n$
$\tilde\chi$ particles. This vertex arises due to electroweak gauge boson
kinetic energy term in $d$ dimensions.}
\label{fig:EW1}
 \begin{minipage}{0.4\textwidth}
 \centering
 \includegraphics[width=0.7\linewidth]{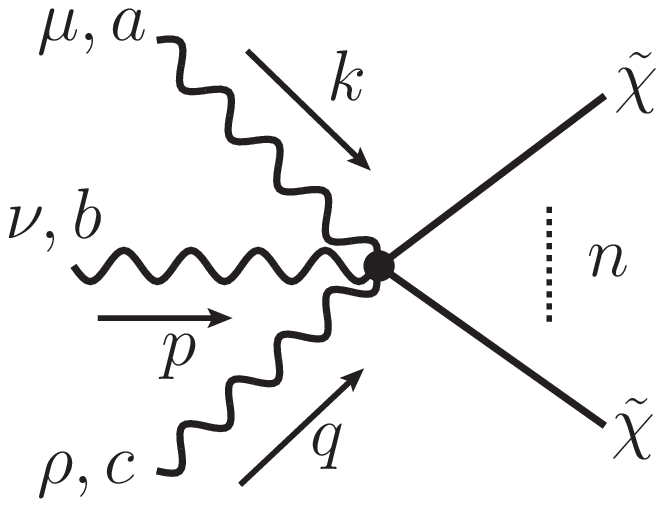}
\end{minipage}
 \begin{minipage}{0.6\textwidth}
 \centering
 \begin{equation*}
\begin{split}
 =\dl(-1)^n &\frac{2 (n-1)!}{\et^{(n-2\dl)}}\,g f^{abc}\\
 &\times \Big(
 g^{\m\n}(k-p)^\rho +  g^{\n\rho}(p-q)^\m +  g^{\rho\m}(q-k)^\n
 \Big)
 \end{split}
\end{equation*}
 \end{minipage}\\
\captionof{figure}{The scattering of three electroweak gauge bosons with $n$
$\tilde\chi$ particles.}
\label{fig:EW2}
   \begin{minipage}{0.4\textwidth}
   \centering
  \includegraphics[width=0.7\linewidth]{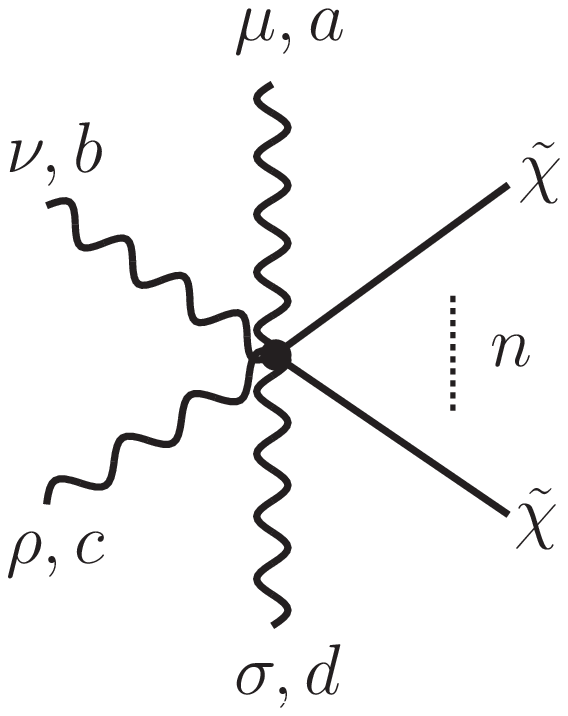} 
   \end{minipage} 
   \begin{minipage}{0.6\textwidth}
   \centering
\begin{equation*}
 \begin{split}
=-i\dl(-1)^n \frac{2 (n-1)!}{\et^{(n-2\dl)}}\times 
 & g^2\Big(
 f^{abe}f^{cde}\left(g^{\n\rho}g^{\n\sigma}-g^{\m\sigma}g^{\n\rho}\right)\\
&+f^{ace}f^{bde}\left(g^{\m\n}g^{\rho\sigma}-g^{\m\sigma}g^{\n\rho}\right)\\
&+f^{ade}f^{bce}\left(g^{\m\n}g^{\rho\sigma}-g^{\m\rho}g^{\n\sigma}\right)
 \Big)
 \end{split}
\end{equation*}
 \end{minipage}\\
 \captionof{figure}{The scattering of four electroweak gauge bosons with $n$
$\tilde\chi$ particles.}
\label{fig:EW3}
%\end{figure}
% \newpage
 \begin{minipage}{0.4\textwidth}
 \centering
  \includegraphics[width=0.7\linewidth]{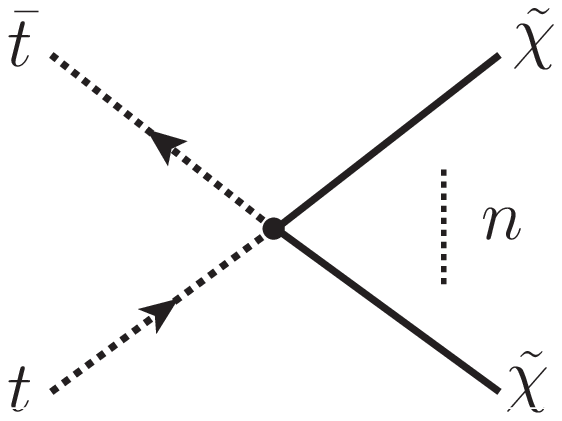} 
   \end{minipage} \hfill
   \begin{minipage}{0.6\textwidth}
   \centering
\bas
=-i\dl(-1)^n(n-1)! \frac{ g_t v}{\et^{n+\dl}}
\eas
 \end{minipage}\\
 \captionof{figure}{The scattering of a top and an anti-top quark with $n$
$\tilde\chi$ particles.}
\label{fig:Yukawa1}

\newpage

\begin{minipage}{0.5\textwidth}
 \includegraphics[width=6.8cm]{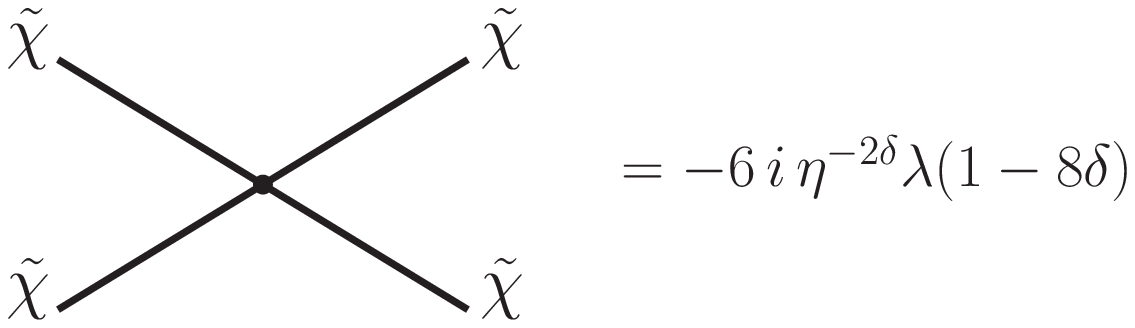} 
 \end{minipage}
\begin{minipage}{0.5\linewidth}
 \includegraphics[width=5.7cm]{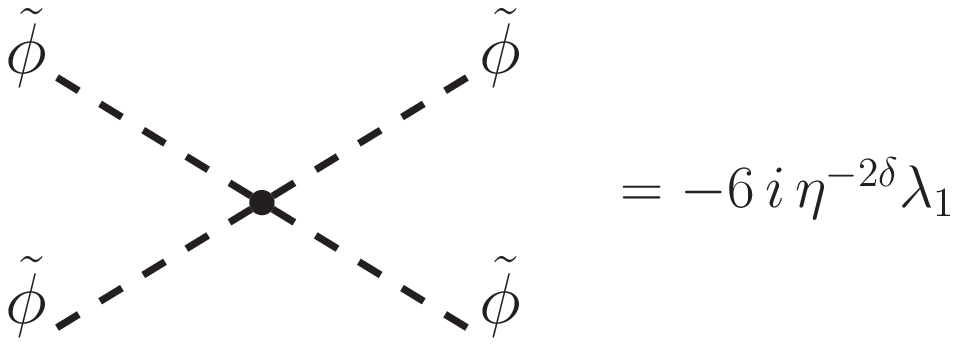}
 \end{minipage}\\
 \begin{minipage}{0.5\textwidth}
 \includegraphics[width=7.3cm]{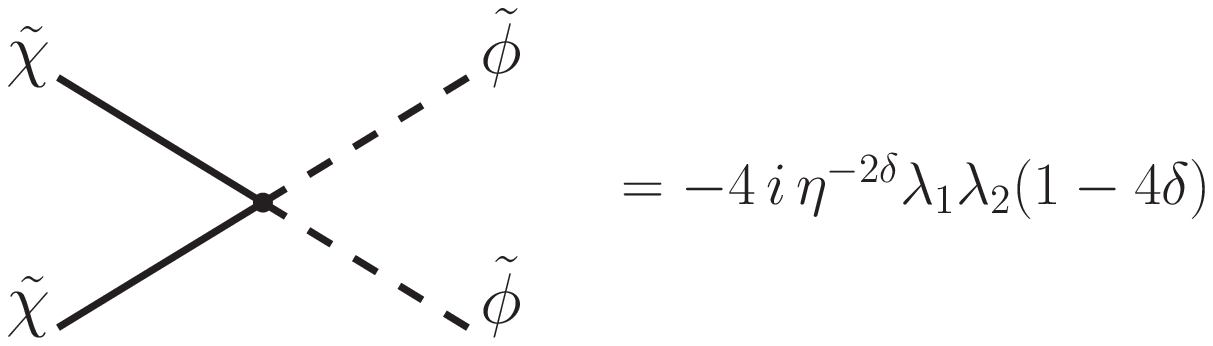}
\end{minipage}
 \begin{minipage}{0.5\linewidth}
\includegraphics[width=7.5cm]{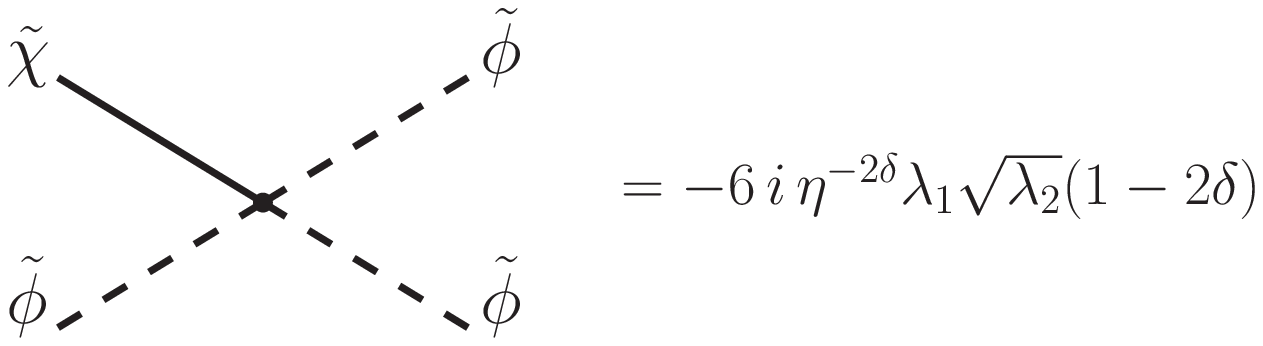}
 \end{minipage}\\
  \begin{minipage}{0.5\textwidth}
  \includegraphics[width=5.9cm]{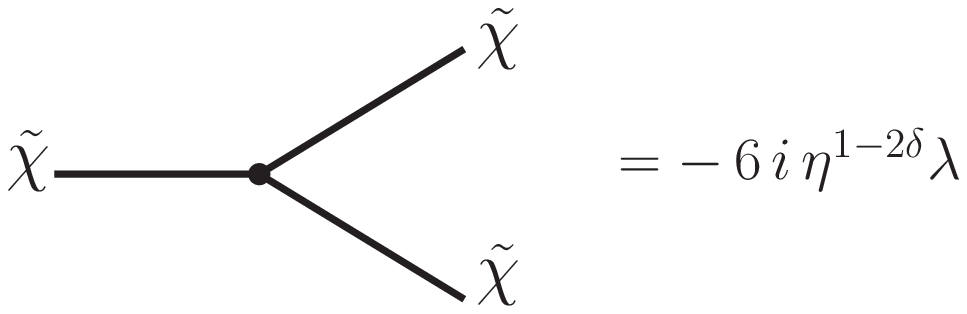} 
   \end{minipage}
\begin{minipage}{0.5\textwidth}
  \includegraphics[width=6.6cm]{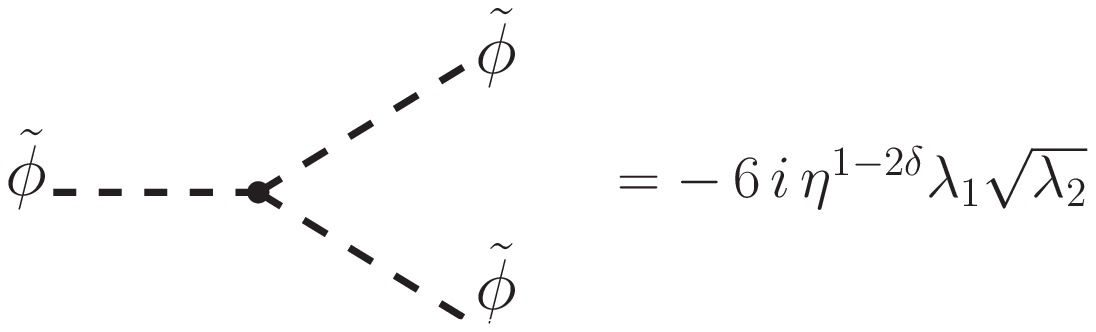} 
   \end{minipage}\\
\begin{minipage}{\textwidth}
  \includegraphics[width=6.2cm]{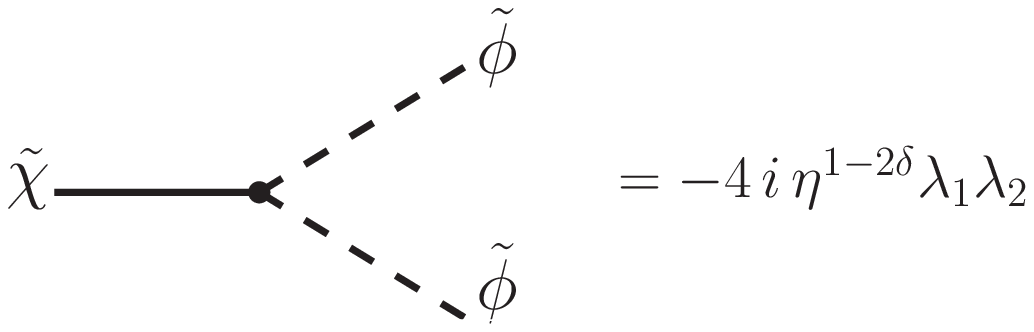} 
   \end{minipage}   
 \captionof{figure}{The Feynman rules arising from the scalar field potential
terms.} 
\label{fig:potential}

\bigskip
\noindent
{\bf \large Acknowledgements:}  Gopal Kashyap thanks the Council
of Scientific and Industrial Research (CSIR), India for providing his Ph.D.
fellowship. PJ thanks Shrihari Gopalakrishna, S. Kalyana
Rama and Romesh Kaul for a stimulating discussion.

\end{spacing}
\begin{spacing}{1}
\begin{small}

\end{small}
\end{spacing}
\end{document}